\documentclass[%
reprint,
 amsmath,amssymb,
 aps,
]{revtex4-2}
\usepackage{float}
\usepackage{graphicx}% Include figure files
\usepackage{dcolumn}% Align table columns on decimal point
\usepackage{bm}% bold math
\usepackage{hyperref}% add hypertext capabilities
\usepackage{lineno}% Enable numbering of text and display math

\begin{document}

%\preprint{APS/123-QED}

\title{Femtoscopic signatures of unique nuclear structures in relativistic collisions}% Force line breaks with \\

\author{D\'aniel Kincses}
 \email{kincses@ttk.elte.hu}
\affiliation{%
 ELTE E\"otv\"os Lor\'and University, P\'azm\'any P\'eter s\'et\'any 1/A, H-1117 Budapest, Hungary
}%


\begin{abstract}
One of the most vital topics of today's high-energy nuclear physics is the investigation of the nuclear structure of the collided nuclei. Recent studies at the Relativistic Heavy Ion Collider (RHIC) and the Large Hadron Collider (LHC) have shown that several observables, such as the collective flow and transverse-momentum correlations of the produced particles, can be sensitive to various nuclear structure and deformation parameters. Femtoscopy, another essential tool for investigating the space-time geometry of the matter created in nuclear collisions, has not yet been widely applied to such studies. Using a multiphase transport model (AMPT), in this Letter, it is demonstrated that the femtoscopic source parameters of pion pairs can also serve as a robust signal of unique nuclear structure. Through an analysis of {$^{208}$Pb{+}$^{20}$Ne} and {$^{208}$Pb{+}$^{16}$O} collisions at ${\sqrt{s_{NN}} = 68.5\textnormal{ GeV}}$, two collision systems especially relevant to the SMOG2 program of the LHCb experiment, it is shown that a deformed initial shape can significantly affect femtoscopic source parameters. This study highlights the importance of expanding the nuclear structure investigations to femtoscopic observables and serves as a baseline for numerous possible future studies in this new direction.
%\begin{description}
%\end{description}
\end{abstract}

%Use showkeys class option if keyword
\maketitle

%\tableofcontents

\section{\label{sec:intro}Introduction}

In recent years, the high-energy nuclear physics community has shown significant interest in the possibility of imaging nuclear structures in high-energy collisions~\cite{Giacalone:2021udy,Bally:2021qys,Giacalone:2023cet,Zhang:2021kxj,Zhao:2024lpc,Giacalone:2023cet,Lacey:2024fpb,Prasad:2024ahm,Magdy:2024thf}. One of the main tools in the arsenal of measurements is the anisotropic flow of particles created in nuclear collisions~\cite{PhysRevD.46.229}. In fluid-dynamical descriptions of nuclear collisions, it has been shown that the momentum anisotropy of the particles originates from the azimuthal anisotropy of the initial density profile of the fireball~\cite{Niemi:2012aj}. Measuring the Fourier coefficients $v_n$ of the single particle azimuthal distributions and investigating their correlations with transverse momentum have become widely used tools for characterizing the parameters of initial-state nuclear deformation~\cite{Giacalone:2020dln,ALICE:2018lao,ALICE:2023tvh,STAR:2024wgy,ATLAS:2022dov,Nielsen:2025pkz,Nielsen:2023znu,Wang:2024vjf}. 

Another important subfield of high-energy physics, femtoscopic correlation measurements~\cite{Lisa:2005dd}, has the potential to become an important asset in this new direction (it has already been utilized to study the nuclear structure in Ref.~\cite{He:2020jzd}). Such measurements provide a highly versatile tool for investigating the space-time geometry of the particle-emitting source created in high-energy nuclear collisions~\cite{Csorgo:1999sj,Wiedemann:1999qn}. At the core of femtoscopy is the equation~\cite{Goldhaber:1960sf,Kopylov:1974th,Podgoretsky:1989bp,Koonin:1977fh,Yano:1978gk,Lednicky:1981su,Pratt:1984su,Pratt:1997pw}:
\begin{align} 
C_2(\vec{q},\vec{K}) = \int d^3\vec{\rho}\, D_{\vec{K}}(\vec{\rho}) |\psi_{\vec{q}}(\vec{\rho})|^2, \label{e:C2D}
\end{align}
which connects the $C_2$ two-particle momentum correlation function to the $D$ spatial correlation function, also known as the pair source function (integrated over the time separation in the pair rest frame). The momentum correlation function depends on the pair relative momentum $\vec{q}$ and the average pair momentum $\vec{K}$, while the pair source function depends on the relative pair separation $\vec{\rho}$. In case of bosonic particles (e.g., pions), the $\Psi_{\vec{q}}(\vec{\rho})$ quantum-mechanical pair wave function is symmetrized, thus in the interaction-free case the momentum correlation will be equal to the Fourier-transform of the pair source function~\cite{Nagy:2023zbg}. In Equation~\ref{e:C2D}, the variables of the pair wave function are taken in the pair rest frame. In experimental analyses, the pair source is often indirectly studied through $C_2$~\cite{PHENIX:2004yan,STAR:2004qya}, or reconstructed via an imaging method~\cite{PHENIX:2007grx,Verde:2001md,Brown:2000aj,Nzabahimana:2023tab}. In event generator models, such as the multi-phase transport model (AMPT), the full phase-space information is available for the created particles, including the freeze-out coordinates. Therefore, the pair source can be directly reconstructed and studied.

Recent experimental measurements~\cite{PHENIX:2024vjp,Kincses:2024sin,PHENIX:2017ino,CMS:2023xyd,NA61SHINE:2023qzr} and phenomenological studies~\cite{Kincses:2024lnv,Csanad:2024jpy,Nagy:2023zbg,Korodi:2022ohn,Kincses:2019rug} showed that the shape of the pion pair-source in high-energy collisions can be described by an elliptically contoured symmetric L\'evy-stable distribution:
\begin{align}
D(\vec{\rho}) = \mathcal{L}(\alpha,R^2,\vec{\rho})=\int \frac{d^3\vec{q}}{(2\pi)^3}\, e^{i\vec{q}\cdot\vec{\rho}} e^{-\frac{1}{2}|\vec{q}^T R^2 \vec{q} |^{\alpha/2}},\label{eq:Levy3Ddef}
\end{align}
where $\alpha$ is called the L\'evy-exponent, characterizing the power-law tail of the source, and $R^2$ is a symmetric $3{\times}3$ matrix, containing the 6 independent L\'evy-scale parameters. These, in the case of Gaussian parameterization of the source function (corresponding to the special $\alpha = 2$ case), are often referred to as the HBT-radii parameters after Hanbury Brown and Twiss, who invented the intensity-interferometry technique in radio astronomy~\cite{HanburyBrown:1954amm,HanburyBrown:1956bqd}. Note that throughout the paper, the elements of the $R^2$ scale parameter matrix are meant as the scale of the pair-source (not that of the single-particle source) and are taken in the Longitudinal Co-Moving System (LCMS)~\cite{Csorgo:1991ej}. The connection between LCMS and the pair rest frame are discussed in detail in Refs.~\cite{Kurgyis:2020vbz,Nagy:2023zbg,Kincses:2024lnv}.

For the sake of simplicity, the experimental measurements are often angle-averaged~\cite{PHENIX:2024vjp,Kincses:2024sin,PHENIX:2017ino,CMS:2023xyd,NA61SHINE:2023qzr}, extracting only a single scale parameter. Three-dimensional investigations~\cite{Kincses:2024lnv, Kurgyis:2018zck} are more complicated, but can reveal further details about the freeze-out source. A common observation of the extracted scale parameters (be it angle-averaged or multi-dimensional) is that they systematically decrease with the average transverse momentum of the pair. This property is often attributed to collective flow~\cite{PhysRevC.84.014908,Csanad:2008af,Cimerman:2017lmm,Lokos:2016fze}. Thus, measurements that are integrated over the azimuthal angle of the pair relative to the reaction plane cannot probe the entire volume of the fireball created in the collision ~\cite{PhysRevC.84.014908}. However, analyses performed relative to the reaction plane~\cite{STAR:2014shf,ALICE:2017gxt,PhysRevC.66.044903,Khyzhniak:2024chj,Csanad:2008af,Cimerman:2017lmm,Lokos:2016fze} add a sensitivity to the shape of the probed fireball part which can be connected to the shape of the entire fireball in a phenomenological Blast Wave Model~\cite{Retiere:2003kf}. A more detailed transport model (AMPT) allows one to connect the probed fireball shape even to the initial conditions related to the structure of the colliding nuclei – the focus of the current manuscript.

Note also that based on hydrodynamic studies~\cite{Csanad:2008af,Cimerman:2017lmm,Lokos:2016fze}, both the elliptic flow and azimuthal femtoscopy can be sensitive to the initial spatial eccentricity and thus the nuclear deformation effects. However, their respective sensitivities can be different, and any of the two contains geometric parameters in an entangled manner. Thus, measuring both elliptic flow and femtoscopic parameters can help in disentangling various effects when determining the initial nuclear geometry.

\section{\label{sec:met}Methods}

To assess the sensitivity of azimuthal femtoscopic measurements to nuclear structure, {$^{208}$Pb{+}$^{20}$Ne} and {$^{208}$Pb{+}$^{16}$O} collisions at ${\sqrt{s_{NN}} = 68.5\textnormal{ GeV}}$ are studied, two systems especially relevant to the SMOG2 fixed-target program of the LHCb experiment~\cite{Lucarelli:2024mes,DiNezza:2024ctw,Hadjidakis:2019vpg}. Using the AMPT parton transport model~\cite{PhysRevC.72.064901}, two different initial nucleon configurations are investigated in each case: the spherically symmetric Woods-Saxon configuration~\cite{dEnterria:2020dwq} and the Nuclear Lattice Effective Field Theory (NLEFT) configuration where nucleons are distributed in a manner that resembles $\alpha-$clusters~\cite{Giacalone:2024luz}. The latter results in a tetrahedron shape for the oxygen nucleus resembling four $\alpha-$clusters, and a bowling pin shape resembling five $\alpha-$clusters for the neon nucleus. The initial nucleon configurations were also rotated randomly. With similar model configurations, it has already been shown that azimuthal anisotropy is sensitive to the deformed initial shape of the neon nucleus in both a transport model~\cite{Lu:2025cni} and a hydrodynamic calculation~\cite{Giacalone:2024ixe}. However, femtoscopic observables have not yet been investigated.

For this study, 100,000 ultra-central ($b=0$) events were simulated for each collision system (Pb+Ne and Pb+O) and for each nucleon configuration (Woods-Saxon and NLEFT). The pseudo-rapidity acceptance of the LHCb experiment is around $2 < \eta < 5$ in the laboratory frame, which roughly corresponds to ${-2.5 < \eta < 0.5}$ in the center-of-mass frame in the case of fixed-target collisions~\cite{Hadjidakis:2019vpg}; thus, only particles within this $\eta$ range were used for the present analysis. To determine the second-order event plane, charged pions, kaons, and protons in the rapidity range of ${-2.5 < \eta < -0.5}$ were used. For the femtoscopic analysis, charged pions were chosen in the kinematic range of ${-0.5 < \eta < 0.5}$ and ${0.2 < p_T\textnormal{ [GeV}/c]< 1.0}$. The second-order event plane angle $\Psi_2$ was calculated similarly to Ref.~\cite{STAR:2014shf}, with the $Q_x$ and $Q_y$ flow-vectors defined as
\begin{align}
    \Psi_{2} &= \frac{1}{2}\arctan\left(\frac{Q_y}{Q_x}\right),
    \label{eq:psi2}\\
    Q_x &= \frac{1}{N}\sum_iw_i\cos(2\phi_i),\\
    Q_y &= \frac{1}{N}\sum_iw_i\sin(2\phi_i),
\end{align}
where $N$ is the total number of particles, the $w_i$ weight is equal to the $p_T$ transverse momentum of the particle, and $\phi$ is the azimuthal angle of the particle.

To construct the pion pair source distribution, same-charge pion pairs were chosen in 5 different ranges of average transverse momentum $k_T$, and 15 different ranges of pair azimuthal angle relative to the second-order event plane. The components of the $D(\vec{\rho})$ source distribution were calculated in the out-side-longitudinal coordinate frame~\cite{Grassberger:1976au,Podgoretsky:1982xu,Bertsch:1993nx,Pratt:1995wm}, where the 'out' direction is along the average transverse momentum of the pair, 'long' is the beam direction, and 'side' is perpendicular to the other two. A boost to LCMS~\cite{Csorgo:1991ej} was also applied (see Equations 13-15 of Ref.~\cite{Kincses:2024lnv}). Following the methodology of Ref.~\cite{Kincses:2024lnv}, one-dimensional projections of the three-dimensional source distribution were constructed along six different directions corresponding to the following unit vectors:
\begin{align}
\begin{array}{l}
\vec e^{(\mathrm{o})} = \big( 1, 0, 0\big),\vspace{2mm}\\
\vec e^{(\mathrm{s})} = \big( 0, 1, 0\big),\vspace{2mm}\\
\vec e^{(\mathrm{l})} = \big( 0, 0, 1\big),
\end{array}\quad
\begin{array}{l}
\vec e^{(\mathrm{os})} = \frac1{\sqrt2}\big( 1, 1, 0\big),\vspace{1mm}\\
\vec e^{(\mathrm{ol})} = \frac1{\sqrt2}\big( 1, 0, 1\big),\vspace{1mm}\\
\vec e^{(\mathrm{sl})} = \frac1{\sqrt2}\big( 0, 1, 1\big).
\end{array}
\label{e:dirs}
\end{align}

Similarly to Ref.~\cite{Kincses:2024lnv}, in the case of a given $k_T$ and ${\varphi_{\rm{pair}}-\Psi_2}$ range, one-dimensional Lévy-stable distributions were fitted simultaneously to the six projections, with seven free parameters: the six independent Lévy-scale parameters of the $R^2$ matrix, and the same Lévy-exponent parameter $\alpha$. The one-dimensional Lévy-stable distributions are defined as~\cite{doi:10.1142/S0217751X25420011}
\begin{align}
    &\mathcal{L}^{1D}(\rho_\nu, \alpha, R_\nu) = \frac{1}{2\pi}\int dq\, e^{iq\rho_\nu} e^{-\frac{1}{2}|qR_\nu|^{\alpha}}, \textnormal{ where}\nonumber\\
    &\rho_\nu = \vec{e}^{(\nu)}\cdot\vec{\rho},\quad R_\nu = \sqrt{\vec{e}^{(\nu),T}R^2\vec{e}^{(\nu)}},\quad\textnormal{and}\nonumber\\
    &\nu = \rm{o,s,l,os,ol,sl}.
    \label{e:1dlevy}
\end{align}

\begin{figure*}
\includegraphics[width=\textwidth]{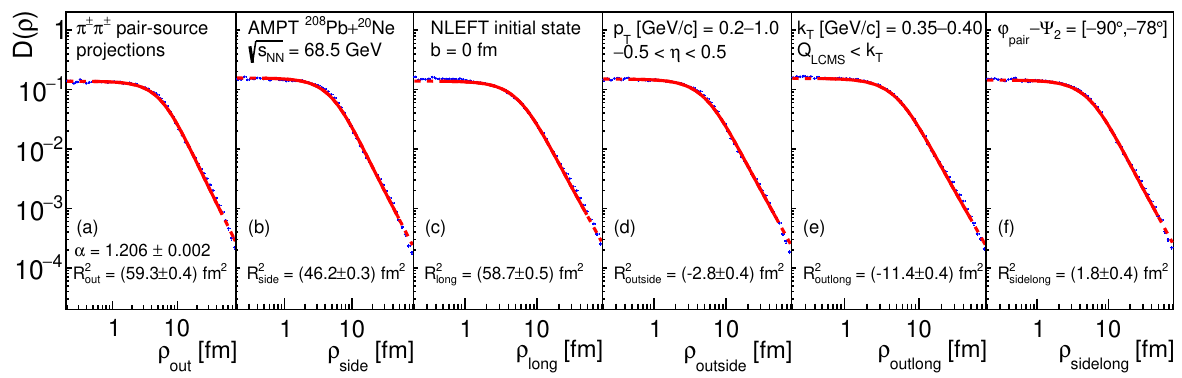}

\caption{\label{fig:sourcefit}An example simultaneous fit to six projections of the three-dimensional source distribution of the same charge pion pairs, reconstructed in AMPT simulations of $\sqrt{s_{NN}}=68.5$ GeV Pb+Ne collisions with the NLEFT initial state nucleon configuration. Panels (a)-(f) show the one-dimensional projections of $D(\vec{\rho})$ with blue markers, corresponding to the directions detailed in Equation~\ref{e:dirs}. The fit with one-dimensional Lévy-stable distributions, as described by Equation~\ref{e:1dlevy}, is shown with red lines.}
\end{figure*}

An example fit is shown in Figure~\ref{fig:sourcefit}.

\begin{figure}
\includegraphics[width=\columnwidth]{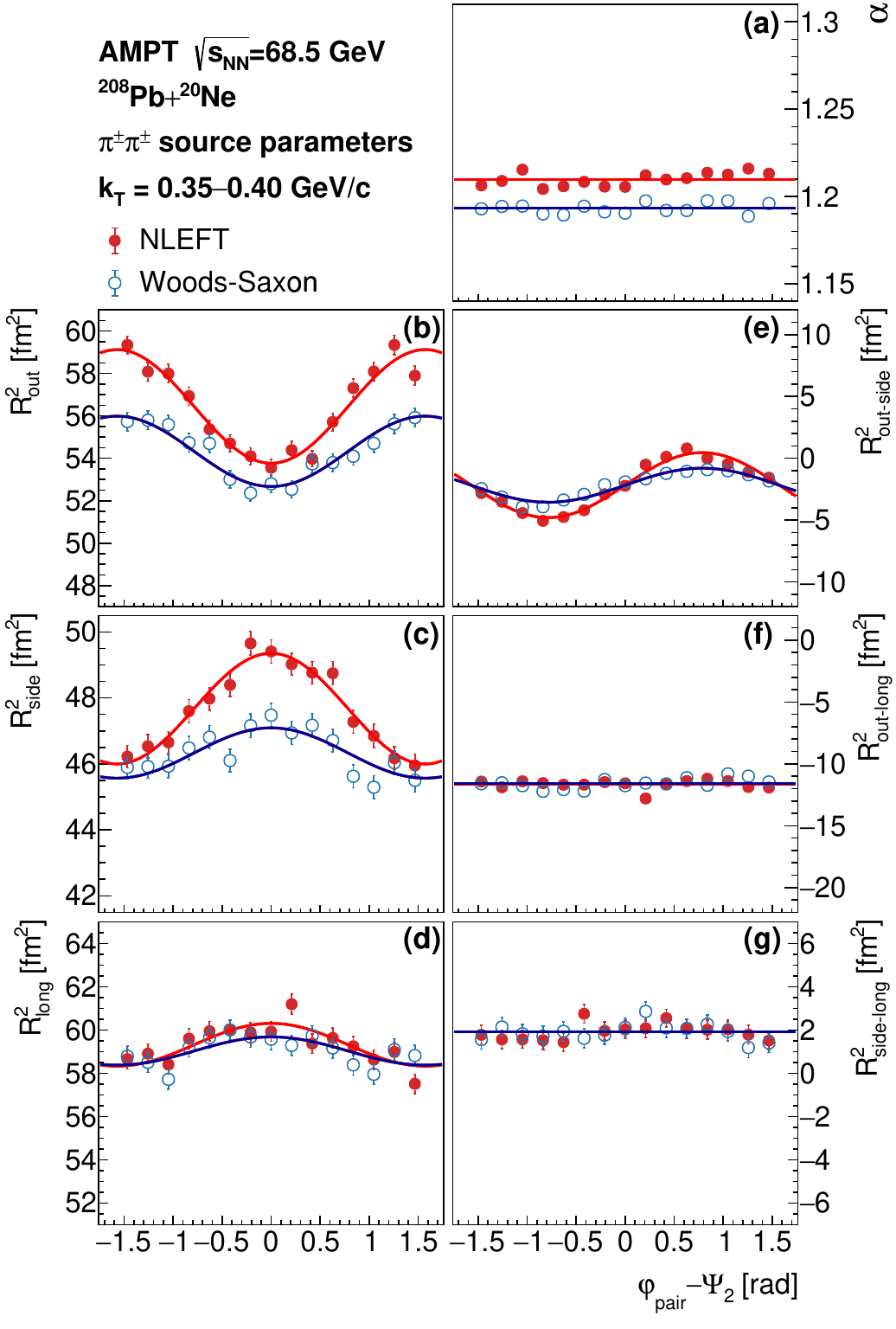}
\caption{\label{fig:sourceparamsPbNe}Extracted pion pair source parameters in ${\sqrt{s_{NN}}=68.5\textnormal{ GeV}}$ ${^{208}\rm{Pb}+^{20}\rm{Ne}}$ collisions generated by AMPT, as a function of pair azimuthal angle relative to the second order event plane, in the average transverse momentum range of ${0.35 < k_T\;(\rm{GeV}/c) < 0.40}$. The source parameter values and their statistical uncertainties are shown with filled red markers and error bars for the NLEFT configuration, and empty blue markers and error bars for the Woods-Saxon configuration. Panel (a) shows the Lévy-exponent parameter $\alpha$, while panels (b)-(g) show the elements of the $R^2$ Lévy-scale parameter matrix. For each dataset, a fit is shown as well corresponding to Equations~\ref{e:Rmu1}~and~\ref{e:Rmu2}.}
\end{figure}

\begin{figure}
\includegraphics[width=\columnwidth]{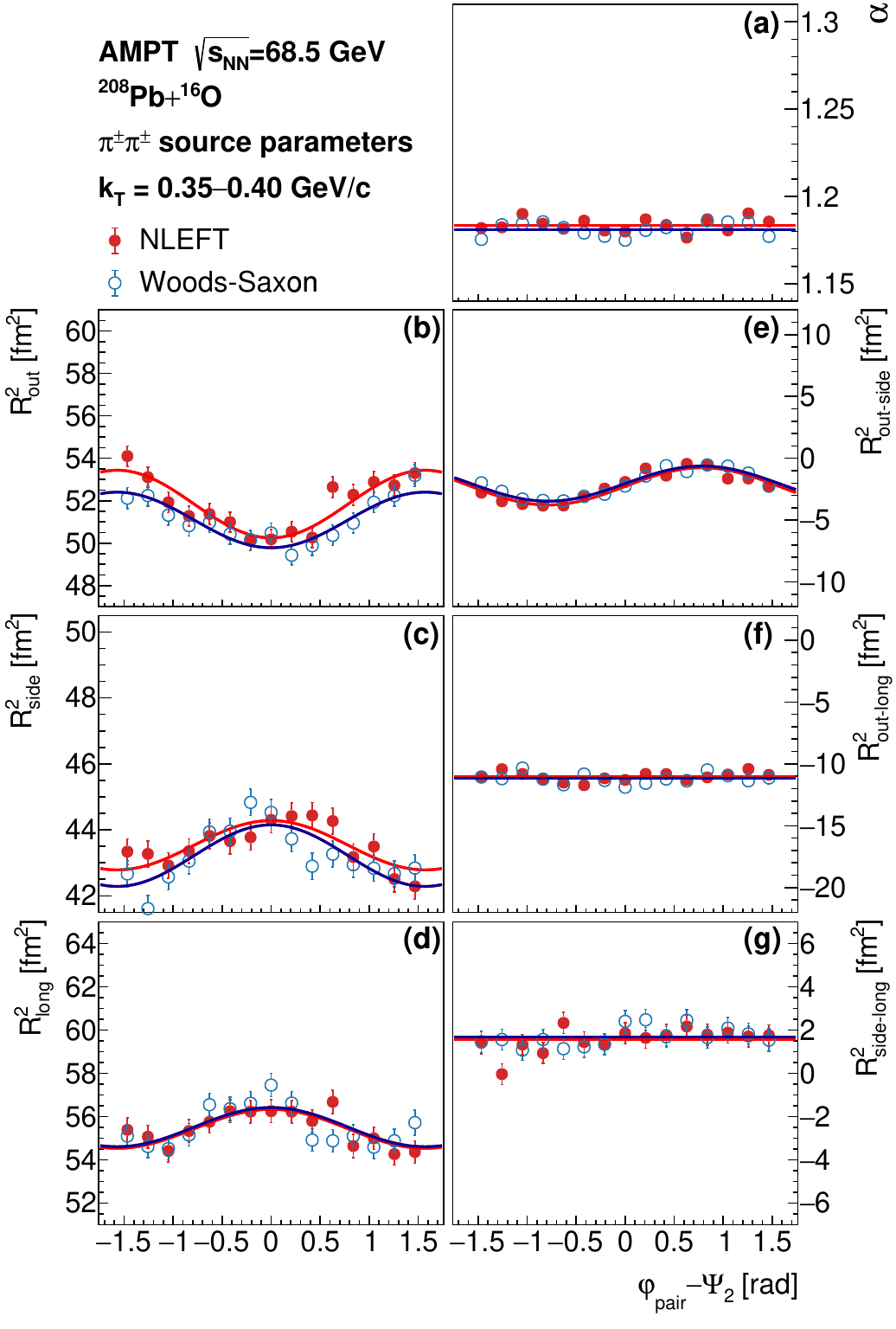}
\caption{\label{fig:sourceparamsPbO}Extracted pion pair source parameters in ${\sqrt{s_{NN}}=68.5\textnormal{ GeV}}$ ${^{208}\rm{Pb}+^{16}\rm{O}}$ collisions generated by AMPT, as a function of pair azimuthal angle relative to the second order event plane, in the average transverse momentum range of ${0.35 < k_T\;(\rm{GeV}/c) < 0.40}$. The source parameter values and their statistical uncertainties are shown with filled red markers and error bars for the NLEFT configuration, and empty blue markers and error bars for the Woods-Saxon configuration. Panel (a) shows the Lévy-exponent parameter $\alpha$, while panels (b)-(g) show the elements of the $R^2$ Lévy-scale parameter matrix. For each dataset, a fit is shown as well corresponding to Equations~\ref{e:Rmu1}~and~\ref{e:Rmu2}.}
\end{figure}

After extracting the source parameters, their $\varphi_{\rm{pair}}-\Psi_2$ dependence were investigated in each $k_T$ bin. For the azimuthal dependence of the scale parameters the following parametrization were used~\cite{STAR:2014shf}:
\begin{align}
    \label{e:Rmu1}&R^2_\mu(\varphi_{\rm{pair}}-\Psi_2) = R^2_{\mu,0}+2R^2_{\mu,2}\cos(2(\varphi_{\rm{pair}}-\Psi_2)),\nonumber\\&\textnormal{if }\mu = \textnormal{out, side, long, out-long}
\end{align}
and
\begin{align}
    \label{e:Rmu2}&R^2_\mu(\varphi_{\rm{pair}}-\Psi_2) = R^2_{\mu,0}+2R^2_{\mu,2}\sin(2(\varphi_{\rm{pair}}-\Psi_2)),\nonumber\\&\textnormal{if }\mu = \textnormal{out-side, side-long}.
\end{align}
An example set of parameters in a given $k_T$ bin, with fits corresponding to Equations~\ref{e:Rmu1}-\ref{e:Rmu2}, is shown in Figure~\ref{fig:sourceparamsPbNe} for $\rm{Pb}+\rm{Ne}$ collisions and in Figure~\ref{fig:sourceparamsPbO} for $\rm{Pb}+\rm{O}$ collisions.

\section{\label{sec:res}Results and discussion}

As illustrated by Figure~\ref{fig:sourcefit}, the three-dimensional elliptically contoured Lévy-stable distribution provides a good description of the source shape, capturing the apparent power-law tail of the pion pair source (which would not be possible with the Gaussian approximation often applied in similar studies~\cite{He:2020jzd}). This is the first investigation where all six independent scale parameters of the Lévy-stable pion pair source are determined, as all previous analyses were azimuthally integrated~\cite{PHENIX:2024vjp,Kincses:2024sin,PHENIX:2017ino,CMS:2023xyd,NA61SHINE:2023qzr,Kincses:2024lnv,Csanad:2024jpy,Nagy:2023zbg,Korodi:2022ohn,Kincses:2019rug}.

Figure~\ref{fig:sourceparamsPbNe} and Figure~\ref{fig:sourceparamsPbO} show the azimuthal angle dependence of the extracted source parameters in $\rm{Pb}+\rm{Ne}$ and $\rm{Pb}+\rm{O}$ collisions, respectively. As expected, the L\'evy scale parameter $\alpha$ shown in panel (a) does not depend strongly on the azimuthal angle, and a constant fit provides a good description. There is a small systematic difference between the Woods-Saxon and NLEFT configurations in the $\rm{Pb}+\rm{Ne}$ case, but it is most probably below the achievable experimental precision. The extracted Lévy-scale parameters are shown in the other panels, fitted with the parametrization described in Equations~\ref{e:Rmu1}-\ref{e:Rmu2}. Panels (b)-(d) show the diagonal elements of the radii matrix, where a clear separation between the Woods-Saxon and the NLEFT configurations can be observed in the out and side directions. Panels (e)-(g) show the off-diagonal elements, where interestingly, each direction exhibits a zeroth-order Fourier term, probably due to the collision-system asymmetry. Some of these parameters already show differences between the two initial nucleon configurations, however, a more robust signal can be expected in the case of relative oscillations. 

\begin{figure}[h]
\includegraphics[width=\columnwidth]{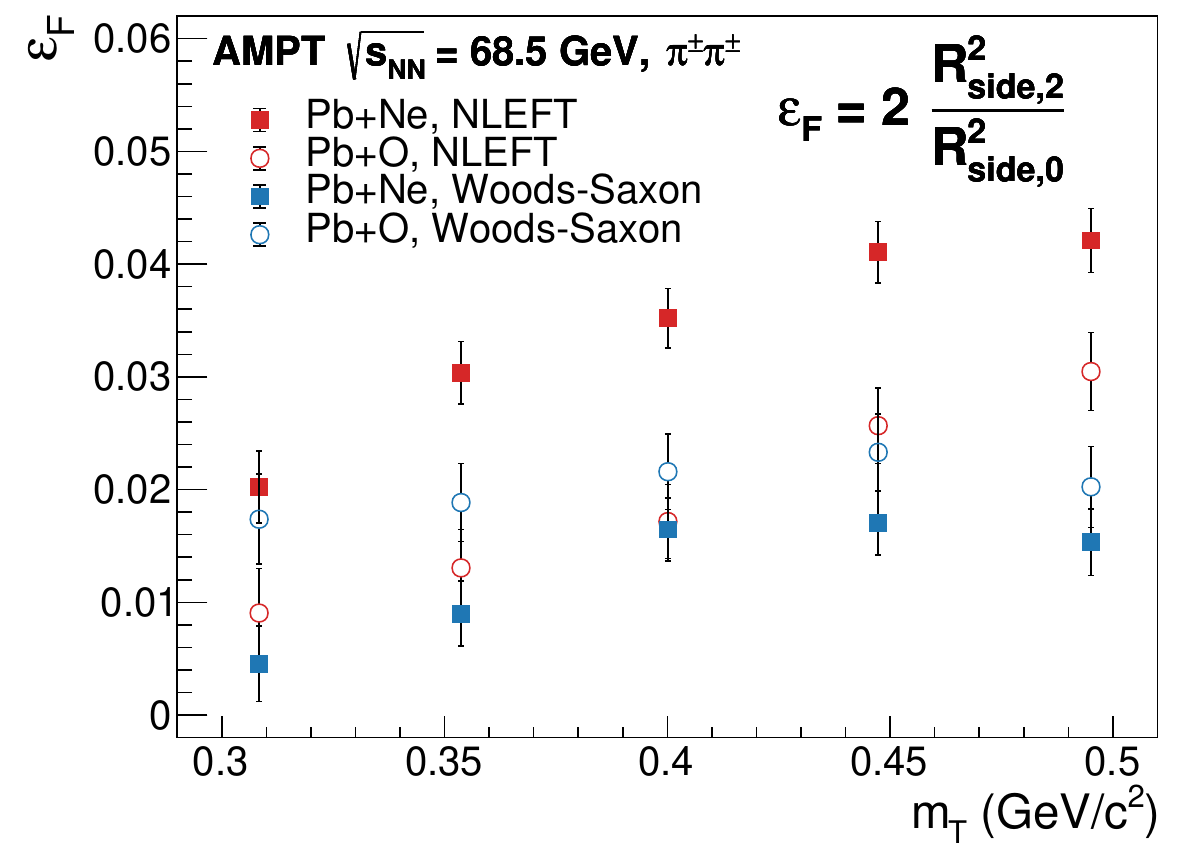}
\caption{\label{fig:ecc}Average transverse mass dependence of the freeze-out eccentricity calculated for four different configurations. The NLEFT and Woods-Saxon initial state configurations are plotted with red and blue markers, respectively. The Pb+Ne result is plotted with filled markers, while the Pb+O result is plotted with empty markers.}
\end{figure}

It has been shown that the relative oscillation (i.e., the ratio of the second to the zeroth order Fourier terms) of the $R^2_{\rm{side}}$ parameter is connected to the freeze-out eccentricity of the fireball around the beam direction~\cite{STAR:2014shf}:
\begin{equation}
    \varepsilon_{F} = 2\frac{R^2_{\rm{side},2}}{R^2_{\rm{side},0}}.
\end{equation}
The dependence of the extracted freeze-out eccentricity on the average transverse mass ${m_T = \sqrt{k_T^2+m_\pi^2}}$ is shown in Figure~\ref{fig:ecc}. For each of the four cases, a slight increase towards higher $m_T$ can be observed. The oxygen results do not exhibit any significant differences between the NLEFT and Woods-Saxon cases, as the clustering-like structure in this case does not increase the elliptical asymmetry. On the other hand, the elliptical asymmetry caused by the bowling-pin shape of the neon nucleus seems to persist through the hadronic phase and significantly increase the freeze-out eccentricity compared to the spherical Woods-Saxon configuration, as well as compared to any of the oxygen configurations. It is also interesting to note that between the two Woods-Saxon cases, the neon eccentricity seems to be systematically below that of the oxygen, probably due to the slightly larger size of the system. 

\begin{figure}[h]
\includegraphics[width=\columnwidth]{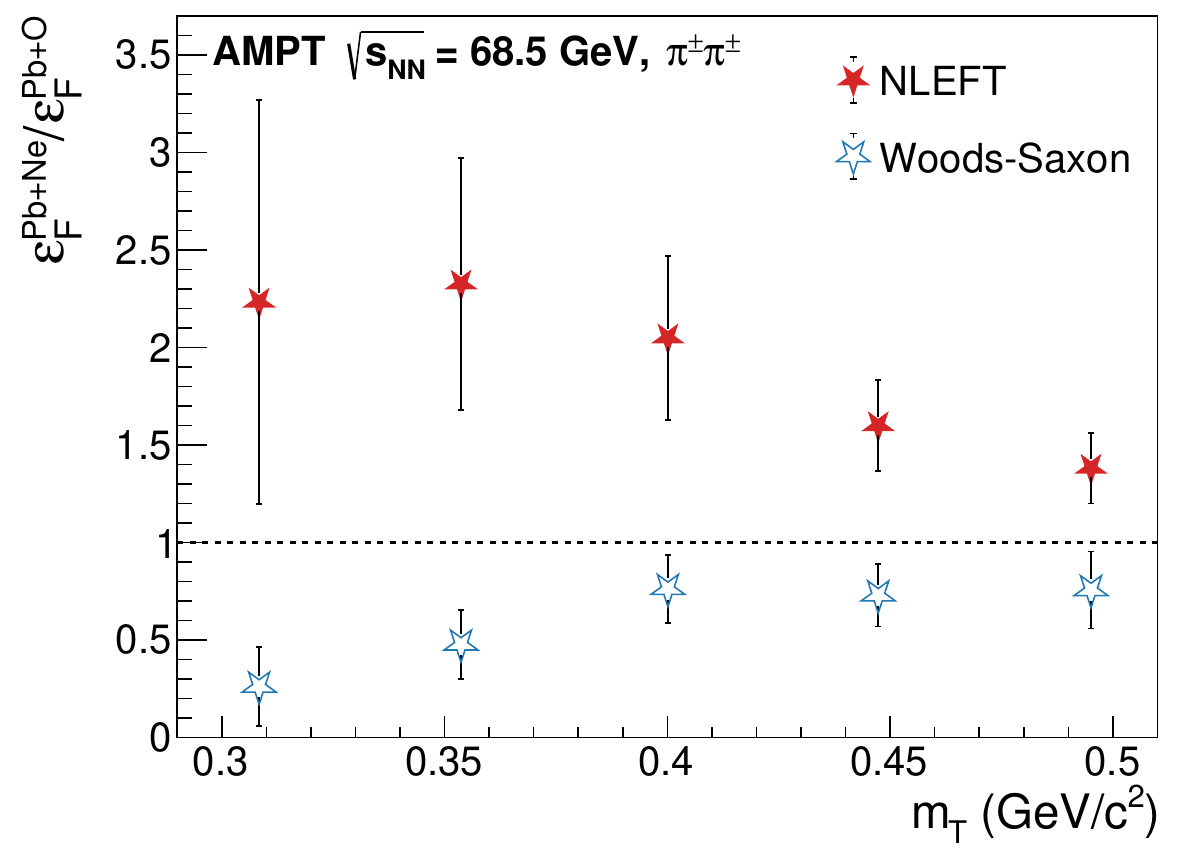}
\caption{\label{fig:ratio}Average transverse mass dependence of the freeze-out eccentricity ratios between Pb+Ne and Pb+O collisions. The NLEFT and Woods-Saxon initial state configurations are plotted with red and blue markers, respectively.}
\end{figure}

In Ref.~\cite{Lu:2025cni}, it has been shown that changing the parton interaction settings within the AMPT model can change the absolute magnitude of the elliptic flow, but the ratio between $\rm{Pb}+\rm{Ne}$ and $\rm{Pb}+\rm{O}$ results remains unaffected. To mitigate systematic uncertainties arising from the evolution of the bulk medium and provide a clearer indication of differences in nuclear structure, observable ratios of the freeze-out eccentricities between $\rm{Pb}+\rm{Ne}$ and $\rm{Pb}+\rm{O}$ collisions are shown in Figure~\ref{fig:ratio}. In the Woods-Saxon case, the ratio is below unity, while for the NLEFT case, it is above unity for the whole $m_T$ range. The difference between the two cases is more pronounced at the lower $m_T$ region. Based on the above observations, the freeze-out eccentricity determined from azimuthally sensitive pion femtoscopy could provide another robust signal for the deformed initial shape of the neon nucleus when compared with oxygen measurements.

\section{\label{sec:sum}Summary and outlook}

This study presents an analysis of $\sqrt{s_{NN}} = 68.5\textnormal{ GeV}$ ${^{208}\rm{Pb}+^{20}\rm{Ne}}$ and ${^{208}\rm{Pb}+^{16}\rm{O}}$ collisions, simulated with the AMPT hadronic transport model. For each collision system, two different initial nucleon configurations are investigated: the spherical Woods-Saxon model and the NLEFT model, where the nucleons are distributed in a manner that resembles $\alpha-$clusters. The pion pair source distribution is investigated in various average transverse mass $m_T$ ranges and pair azimuthal angle ranges relative to the second-order event plane. It is shown that a three-dimensional Lévy-stable distribution provides a good approximation for the source shape, and the Lévy-exponent $\alpha$ and $R^2$ Lévy-scale matrix parameters are determined. Subsequently, from the azimuthal oscillation of the $R^2_{\rm{side}}$ scale parameter, the zeroth- and second-order Fourier components are determined, and from their ratio, the freeze-out eccentricity is calculated. This observable is found to be significantly increased in the NLEFT configuration of Pb+Ne collisions compared to the other three cases. When comparing the Woods-Saxon configurations, it is found that the Pb+Ne eccentricity is systematically below the Pb+O result. Thus, the freeze-out eccentricity determined simultaneously in Pb+Ne and Pb+O collisions from azimuthally sensitive pion femtoscopy could provide another robust signal for the deformed initial shape of the neon nucleus.

Femtoscopy is a rich field with many more directions to explore regarding possible future nuclear structure studies. The azimuthally sensitive analysis could be extended to non-identical particle femtoscopy, as well as to higher-order event planes where similar relative oscillations of the scale parameters might provide more insight. It is important furthermore to extend the investigations to hydrodynamic and hybrid models as well (including both hydrodynamic evolution and hadronic transport).

\begin{acknowledgments}
The author is grateful to You Zhou, M\'at\'e Csan\'ad, and M\'arton Nagy for their feedback and discussions. 
This research was funded by the NKFIH grants TKP2021-
NKTA-64, PD-146589, and K-138136. D. K. was supported by the EK\"OP-24 University Excellence Scholarship program of the Ministry for Culture and Innovation from the source of the national research, development, and innovation fund.
\end{acknowledgments}

\newpage
%\nocite{*}

\bibliography{ref}% Produces the bibliography via BibTeX.

\end{document}